\documentstyle[pra,aps,psfig]{revtex}

\begin{document}
\bibliographystyle{prsty}
\draft

\title{\bf Implementation of quantum gates and preparation of entangled
states in cavity QED with cold trapped ions} 
\author{Mang Feng$^{1,2}$ 
\thanks{Electronic address: feng@isiosf.isi.it},
 Xiaoguang Wang$^{1}$
\thanks{Electronic address: xgwang@isiosf.isi.it}} 
\address{$^{1}$ Institute for Scientific Interchange (ISI) Foundation,\\ 
Villa Gualino, Viale Settimio Severo 65, I-10133, Torino, Italy\\
$^{2}$Laboratory of Magnetic Resonance and Atomic and Molecular Physics,\\ 
Wuhan Institute of Physics and Mathematics, Academia Sinica, Wuhan, 430071, China}
\date{\today}
\maketitle

\begin{abstract}

We propose a scheme to perform basic gates of quantum computing and prepare entangled states in a system   
with cold trapped ions located in a single mode optical cavity.
General quantum computing can be made with both motional state of the trapped ion and cavity state 
being qubits. We can also generate different kinds of entangled states in such a system without state 
reduction, and can transfer quantum states from the ion in one trap to the 
ion in another trap. Experimental requirement for achieving our scheme is discussed. 

\end{abstract}
\vskip 0.1cm
\pacs{PACS numbers: 03.67.-a, 32.80.Lg. 42.50.-p}

\section {introduction}

Both cavity QED and cold trapped ion systems have been drawn much attention over past decades, 
due to not only the fundamental interest of physics involved but also some potential applications, such as
quantum information processing [1]. To our knowledge, the entangled states
between atoms and cavity [2], between ions' internal states and motional
states [3], and between internal states of different ions [4] have been experimentally
achieved. If we combine the cavity QED technique with ion trap one, for
example, placing the ion trap into an optical cavity, the problem would be
more interesting because it involves $three~quantum~degrees~of~freedom$, namely,
the ion's internal levels, phonon field of the trap and photon field of the
cavity [5]. It is of great importance in the field of quantum information to
study entanglement of three quantum degrees of freedom as well as information
transfer between one of them to another. In early papers, Zeng and Lin
proposed a technique to generate nonclassical vibrational states of cold
trapped ions in a harmonic trap located in an optical cavity [6]. Buzek 
$et~al$ [7] and Luo $et~al$ [8] investigated independently the quantum
motion of a cold trapped ion interacting with a quantized
light field. Parkins and his cooperators [9] also studied the coherent transfer
between motional states and quantized light in detail. More recently, we
notice that a work under similar consideration emphasized the production of
Bell states between motional states of a trapped ion and the quantized
light [10]. If we consider entanglement to be important source of quantum
information processing, all the works referred to above have clearly shown
 that this kind of source does exist in the ion-trap-cavity system, whereas 
none of them showed specifically how to use this source for  processing
quantum information. 

Various proposals have been put forward by using internal states of cold atoms in cavities or in ion traps
to be qubits for performing quantum computing and quantum communication [11]. In some proposals, motional 
states of the trapped ion are employed to be qubits [12]. Recently, two proposals [13,14] were presented, 
which involve the ion-trap-cavity system for robust quantum computing against decoherence by 
means of large detuning between the
cavity state and internal states of trapped ions. In fact, if the cavity decay can be effectively used, 
entangled states can be prepared in the cavity [15] and the 
cavity system would become a node of quantum communication network [16,17]. But in the present work, 
we will only consider an $ideal$ case without decoherence, and focus on the construction of controlled-NOT 
(CNOT) and Hadamard gates in the system with a single ion confined in the trap which is by its turn 
located in a high-Q cavity [5], by $employing$ 
$the~motional~state~of~the~ion~and~the~cavity~state~to~be~qubits$. To show the usefulness of the quantum 
gates we present, our model would be extended to two traps,
with the single ion confined in each trap, placed in the same high-Q cavity. As each trapped ion is coupled
to the same cavity state, we can 
prepare entanglement and transfer states between these two subsystems of trapped ions. The experimental 
aspect for achieving our scheme would be discussed.

\section {Quantum computing in the system of a single trapped ion in the high-Q cavity}

Consider a single cold ion confined in the ion trap which itself is located in an optical 
cavity. We assume that the cavity mode, together with classical standing waves of lasers, couples to 
the internal and vibrational states of the ion, as discussed in Refs.[7,8,10]. The Hamiltonian in
the unit of $\hbar=1$ can be written as follows
\begin{equation}
H= \frac {\omega_{0}}{2}\sigma_{z}+\nu b^{\dagger}b +\omega_{c} a^{\dagger}a +
G[\sigma_{+}e^{i[\eta_{L}(b^{\dagger}+b)-\omega_{L}t]}+ h.c]+ g\sigma_{x}(a^{\dagger}+ a)
\sin[\eta_{c}(b^{\dagger}+b)+\phi] 
\end{equation}
where $\omega_{0}$ is the frequency of atomic resonance transition. $\omega_{c}$ and $\omega_{L}$ are 
frequencies of cavity mode and laser respectively. $a^{\dagger}$, $a$ and $b^{\dagger}$, $b$ are respectively
creation and annihilation operators of photons of the cavity and phonons of
the trap. $G$ and $g$ are the coupling constants proportional to the ion-laser
and ion-cavity interaction respectively. $\eta_{L}$ and $\eta_{c}$ are respectively Lamb-Dicke parameters
with respect to the radiation of laser and cavity. $\sigma_{+}$, $\sigma_{-}$ and $\sigma_{z}$ are usual 
Pauli operators. $\phi$ accounts for the  relative position of motional state of the
ion to the standing wave of the quantized cavity field. Within the  Lamb-Dicke
limit, $\sin[\eta_{c} (b^{\dagger}+b)+\phi]\approx \eta_{c} (b^{\dagger}+b)\cos\phi +\sin\phi$.
Performing a unitary  operator defined as  $U=\exp [-it(\frac
{\omega_{0}}{2}\sigma_{z}+\nu b^{\dagger}b +\omega_{c} a^{\dagger}a) ]$
yields different cases under the rotating wave approximation, for example, \\
case 1 with $\omega_{L}=\omega_{0}$,~~~~~~~~~~~~~~~~~ $H^{I}_{1}= G (\sigma_{+}+ \sigma_{-});$\\
case 2 with $\omega_{L}=\omega_{0}-\nu$,~~~~~~~~~~~ $H^{I}_{2}=iW (b\sigma_{+}-b^{\dagger}\sigma_{-});$\\
case 3 with $\omega_{L}=\omega_{0}+\nu$,~~~~~~~~~~~ $H^{I}_{3}=iW (b^{\dagger}\sigma_{+}-b\sigma_{-});$\\
case 4 with $\omega_{c}=\omega_{0}-\nu$,~~~~~~~~~~~ $H^{I}_{4}=\Omega (a^{\dagger}b^{\dagger}\sigma_{-}+ ab\sigma_{+});
$\\
case 5 with $\omega_{c}=\nu- \omega_{0}$,~~~~~~~~~~~ $H^{I}_{5}=\Omega (ab^{\dagger}\sigma_{-}+ a^{\dagger}b\sigma_{+});
$\\
case 6 with $\omega_{c}=\omega_{0}+\nu$,~~~~~~~~~~~ $H^{I}_{6}=\Omega (ab^{\dagger}\sigma_{+}+
a^{\dagger}b\sigma_{-});
$\\
case 7 with $\omega_{c}=\omega_{0}$,~~~~~~~~~~~~~~~~ $H^{I}_{7}=\Omega' (a^{\dagger}\sigma_{-}+ a\sigma_{+}),$\\
where $W=G\eta_{L}$, $\Omega=\eta_{c}g\cos\phi$ and $\Omega'=g\sin\phi$. 

Therefore by adjusting the laser or cavity frequency, we can obtain different time evolutions of states in the system. In
what follows, we only list some of them which would be used for constructing quantum computing gates and
preparing entangled states in the present work. \\
For case 1, 
\begin{equation}
|g\rangle \rightarrow \cos(G t) |g\rangle -i\sin (G t) |e\rangle,~~|e\rangle \rightarrow \cos(G t) |e\rangle -i\sin (G t) |g\rangle;
\end{equation}
for case 2,
$$|g\rangle|m\rangle_{b} \rightarrow \cos(\sqrt{m}W t) |g\rangle|m\rangle_{b} +\sin (\sqrt{m}W t)
|e\rangle|m-1\rangle_{b},$$ 
\begin{equation}
|e\rangle|m\rangle_{b} \rightarrow \cos(\sqrt{m+1}W t) |e\rangle|m\rangle_{b} -\sin (\sqrt{m+1}W t) |g
\rangle|m+1\rangle_{b};
\end{equation}
for case 4,\\
$$|g\rangle|m\rangle_{a}|n\rangle_{b} \rightarrow \cos(\sqrt{mn}\Omega t) |g\rangle|m\rangle_{a}|n\rangle_{b} 
-i\sin (\sqrt{mn}\Omega t) |e\rangle|m-1\rangle_{a}|n-1\rangle_{b},$$ 
\begin{equation}
|e\rangle|m\rangle_{a}|n\rangle_{b} \rightarrow \cos(\sqrt{(m+1)(n+1)}\Omega t)|e\rangle|m\rangle_{a}|n\rangle_{b}
-i\sin (\sqrt{(m+1)(n+1)}\Omega t) |g\rangle|m+1\rangle_{a}|n+1\rangle_{b};
\end{equation}
for case 6,\\
$$|g\rangle|m\rangle_{a}|n\rangle_{b} \rightarrow \cos(\sqrt{m(n+1)}\Omega t) |g\rangle|m\rangle_{a}|n\rangle_{b} 
-i\sin (\sqrt{m(n+1)}\Omega t) |e\rangle|m-1\rangle_{a}|n+1\rangle_{b},$$ 
\begin{equation}
|e\rangle|m\rangle_{a}|n\rangle_{b} \rightarrow \cos(\sqrt{(m+1)n}\Omega t) |e\rangle|m\rangle_{a}|n\rangle_{b} 
-i\sin (\sqrt{(m+1)n}\Omega t) |g\rangle|m+1\rangle_{a}|n-1\rangle_{b};
\end{equation}
for case 7,\\
$$|g\rangle|m\rangle_{a} \rightarrow \cos(\sqrt{m}\Omega't) |g\rangle|m\rangle_{a} -i\sin (\sqrt{m}\Omega't)
|e\rangle|m-1\rangle_{a},$$ 
\begin{equation}
|e\rangle|m\rangle_{a} \rightarrow \cos(\sqrt{m+1}\Omega't) |e\rangle|m\rangle_{a} -i\sin (\sqrt{m+1}\Omega't) 
|g\rangle|m+1\rangle_{a}
\end{equation}
where the subscripts $a$ and $b$ denote the states of photon and phonon respectively. $|g\rangle$ and $|e\rangle$ are 
ground and excited internal states of the ion respectively, and $m,n= 0,1,\cdots$.  

With above time evolutions of states for different cases, we can construct Hadamard and
CNOT gates in the system, which are basic elements of a general quantum computing [18]. 
We first try to implement CNOT gate between $|\rangle_{a}$ and $|\rangle_{b}$.
If the internal state of the ion is prepared to the ground state, by performing operations 
$R^{I}_{4}(\frac {\pi}{2})$, $R^{I}_{7}(\frac {3\pi}{2})$ and $R^{I}_{4}(\frac {\pi}{2})$ sequentially, where 
$R^{I}_{4}(\frac {\pi}{2})$ means the time evolution of the case 4 with $\Omega t=\frac {\pi}{2}$ and 
$R^{I}_{7}(\frac {3\pi}{2})$ is the time evolution of the case 7 with 
$\Omega 't=\frac {3\pi}{2}$, we have
$$|g\rangle|10\rangle_{ab}\rightarrow |g\rangle|10\rangle_{ab}\rightarrow i|e\rangle|00\rangle_{ab}\rightarrow |g\rangle|11\rangle_{ab},$$
$$|g\rangle|01\rangle_{ab}\rightarrow |g\rangle|01\rangle_{ab}\rightarrow |g\rangle|01\rangle_{ab}\rightarrow |g\rangle|01\rangle_{ab},$$
$$|g\rangle|11\rangle_{ab}\rightarrow -i|e\rangle|00\rangle_{ab}\rightarrow |g\rangle|10\rangle_{ab}\rightarrow |g\rangle|10\rangle_{ab},$$ 
\begin{equation}
|g\rangle|00\rangle_{ab}\rightarrow |g\rangle|00\rangle_{ab}\rightarrow |g\rangle|00\rangle_{ab}\rightarrow |g\rangle|00\rangle_{ab}
\end{equation}
which is the CNOT gate with $|\rangle_{a}$ and $|\rangle_{b}$ being control and target states respectively. 
Similarly, to perform the CNOT gate with $|\rangle_{b}$ and $|\rangle_{a}$ being control and target states 
respectively, we combine $H^{I}_{4}$ and $H^{I}_{2}$. With operations of $R^{I}_{4}(\frac {3\pi}{2})$, 
$R^{I}_{2}(\frac {3\pi}{2})$ and  $R^{I}_{4}(\frac {\pi}{2})$ sequentially, we have
$$|g\rangle|10\rangle_{ab}\rightarrow |g\rangle|10\rangle_{ab}\rightarrow |g\rangle|10\rangle_{ab}
\rightarrow |g\rangle|10\rangle_{ab},$$
$$|g\rangle|01\rangle_{ab}\rightarrow |g\rangle|01\rangle_{ab}\rightarrow -|e\rangle|00\rangle_{ab}
\rightarrow i|g\rangle|11\rangle_{ab},$$
$$|g\rangle|11\rangle_{ab}\rightarrow i|e\rangle|00\rangle_{ab}\rightarrow i|g\rangle|01\rangle_{ab}
\rightarrow i|g\rangle|01\rangle_{ab},$$ 
\begin{equation}
|g\rangle|00\rangle_{ab}\rightarrow |g\rangle|00\rangle_{ab}\rightarrow |g\rangle|00\rangle_{ab}\rightarrow
|g\rangle|00\rangle_{ab}.
\end{equation}
So further single qubit rotation is needed for eliminating the prefactor $i$ when the final result of motional
state of ion is $|1>_{b}$.

For implementing a Hadamard gate of $|\rangle_{a}$, we 
also need to initially prepare the internal state of the ion to the ground state. By performing operations 
$R^{I}_{7}(\frac {\pi}{2})$, $R^{I}_{1}(\frac {7\pi}{4})$ and
$R^{I}_{7}(\frac {\pi}{2})$, we obtain 
$$|g\rangle|1\rangle_{a}\rightarrow
-i|e\rangle|0\rangle_{a}\rightarrow
\frac {-1}{\sqrt{2}}i(|e\rangle+i|g\rangle)|0\rangle_{a}\rightarrow
\frac {1}{\sqrt{2}}|g\rangle(|0\rangle_{a}-|1\rangle_{a}),$$ 
\begin{equation}
|g\rangle|0\rangle_{a}\rightarrow |g\rangle|0\rangle_{a}\rightarrow
\frac {1}{\sqrt{2}}(|g\rangle+i|e\rangle)|0\rangle_{a}\rightarrow
\frac {1}{\sqrt{2}}|g\rangle(|0\rangle_{a}+|1\rangle_{a}).
\end{equation} 
 The Hadamard gate of $|\rangle_{b}$ can be obtained similarly by replacing $R^{I}_{7}(\frac
{\pi}{2})$ with $R^{I}_{2}(\frac {\pi}{2})$.  

Therefore, the general quantum computing can be made with the bosonic qubits, i.e., the motional state of the
trapped ion and the cavity state being qubits. However, it would be more meaningful if we consider a system 
with more trapped ions in a cavity. Therefore in next section, we will discuss 
 how to use above quantum computing gates to transfer states and prepare entangled states in the case that
two trapped ions are coupled to the same cavity state. 

\section {Quantum communication in the system of  two trapped ions in A high-Q cavity}

Suppose that there are two separate traps placed in the same high-Q optical cavity with a single ion confined
in each trap. If Coulomb interaction between the two ions is weak enough, and the two subsystems of trapped 
ions are identical, we can obtain Eqs.(2)-(9) for the two subsystems respectively, where the cavity state 
$|\rangle_{a}$ plays the role of connection between these two subsystems. Although we only focused our 
attention on
the two bosonic qubits in last section, we will investigate in this section how to make entangled states and 
state transfer with all quantum degrees of freedom involved. 

Application 1: State transfer between two trapped ions 

Suppose that the two trapped ions are respectively prepared to $C|g\rangle_{1}+D|e\rangle_{1}$ and $|g\rangle_{2}$
with $|C|^{2}+|D|^{2}=1$ and subscripts being labeling of the ions. Both 
motional states of the two ions and the cavity state are prepared to ground
states. So we have $(C|g\rangle_{1}+D|e\rangle_{1})|0\rangle_{b_{1}}|g\rangle_{2}|0\rangle_{b_{2}}|0\rangle_{a}$ with
subscripts $b_{1}$ and $b_{2}$ labeling motional states of ions 1 and 2
respectively. By performing operations  $R^{I_{1}}_{7}(\frac {\pi}{2})$ and then
$R^{I_{2}}_{7}(\frac {3\pi}{2})$, where $R^{I_{1}}_{7}$ and $R^{I_{2}}_{7}$ are
time evolutions of the case 7 for ions 1 and 2 respectively, we have  
$$ (C|g\rangle+D|e\rangle)_{1}|0\rangle_{b_{1}}|g\rangle_{2}|0\rangle_{b_{2}}|0\rangle_{a} \rightarrow
|g\rangle_{1}|0\rangle_{b_{1}}|g\rangle_{2}|0\rangle_{b_{2}}(C|0\rangle-iD|1\rangle)_{a} $$
\begin{equation}
\rightarrow |g\rangle_{1}|0\rangle_{b_{1}}(C|g\rangle+D|e\rangle)_{2}|0\rangle_{b_{2}}|0\rangle_{a}. 
\end{equation}
The state transfer shown above is made through coupling to the cavity state, instead of the direct 
interaction 
between the two trapped ions. So if there are many trapped ions placed in the cavity, we can transfer the
state from one trapped ion to another even if they are not neighboring. In fact, we can also perform swapping
operation between above two trapped ions. To this end, we have to define 
$S_{ab}=CNOT_{ab}$ $\cdot$ $CNOT_{ba}$ $\cdot$ $CNOT_{ab}$ where $S_{ab}$ is the operation of state 
exchange between $|\rangle_{a}$ and 
$|\rangle_{b}$, and $CNOT_{ij}$ the CNOT gate with $|\rangle_{i}$ and $|\rangle_{j}$ being control and target states
respectively. If we start our process from the initial state of 
$(C|g\rangle+D|e\rangle)_{1}|0\rangle_{b_{1}}(E|g\rangle+F|e\rangle)_{2}|0\rangle_{b_{2}}|0\rangle_{a}$ 
where $|C|^{2}+|D|^{2}=1$ and $|E|^{2}+|F|^{2}=1$, performing 
$R^{I_{1}}_{2}(\frac {\pi}{2})$, $R^{I_{2}}_{2}(\frac {\pi}{2})$, $S_{ab_{1}}, S_{ab_{2}}, S_{ab_{1}}$,
$R^{I_{1}}_{2}(\frac {3\pi}{2})$, and $R^{I_{2}}_{2}(\frac {3\pi}{2})$ sequentially will yield
$$ (C|g\rangle+D|e\rangle)_{1}|0\rangle_{b_{1}}(E|g\rangle+F|e\rangle)_{2}|0\rangle_{b_{2}}|0\rangle_{a} \rightarrow
|g\rangle_{1}(C|0\rangle-D|1\rangle)_{b_{1}}|g\rangle_{2}(E|0\rangle-F|1\rangle)_{b_{2}}|0\rangle_{a} 
\rightarrow$$
\begin{equation}
|g\rangle_{1}(E|0\rangle-F|1\rangle)_{b_{1}}|g\rangle_{2}(C|0\rangle-D|1\rangle)_{b_{2}}|0\rangle_{a}
\rightarrow (E|g\rangle+F|e\rangle)_{1}|0\rangle_{b_{1}}(C|g\rangle+D|e\rangle)_{2}|0\rangle_{b_{2}}|0
\rangle_{a}. 
\end{equation}
Briefly speaking, the above swapping involves three steps, i.e., quantum
information transferring from ions'
internal states to corresponding motional states, then swapping between motional states of the two ions, and 
then quantum information transferring from ions' motional states to corresponding internal states.

Application 2: Preparation of Greenberger-Horne-Zeilinger (GHZ) state [20]

Suppose that the initial state of the system is the same as that of Eq.(10).  By first performing operation 
$R^{I_{1}}_{7}(\frac {\pi}{2})$, then CNOT$_{ab_{1}}$ and CNOT$_{ab_{2}}$ respectively, we have
$$(C|g\rangle+D|e\rangle)_{1}|0\rangle_{b_{1}}|g\rangle_{2}|0\rangle_{b_{2}}|0\rangle_{a} \rightarrow 
|g\rangle_{1}|0\rangle_{b_{1}}|g\rangle_{2}|0\rangle_{b_{2}} (C|0\rangle-iD|1\rangle)_{a}$$
\begin{equation}
\rightarrow |g\rangle_{1}|g\rangle_{2}(C|000\rangle-iD|111\rangle)_{b_{1}b_{2}a}.
\end{equation}
The GHZ state prepared here is different from that in Ref.[7]. It is more interesting because we can control 
the entanglement of the GHZ state by preparing the initial state of the ion 1. If many trapped ions are
located in the same cavity, we can obtain a GHZ state with many qubits along this idea.

Application 3: Entanglement of the two trapped ions

After obtaining the GHZ state presented in Eq.(12), we perform a CNOT$_{b_{1}a}$. Then we have
\begin{equation}
 |g\rangle_{1}|g\rangle_{2}(C|000\rangle-iD|111\rangle)_{b_{1}b_{2}a} \rightarrow 
|g\rangle_{1}|g\rangle_{2}(C|00\rangle-iD|11\rangle)_{b_{1}b_{2}}|0\rangle_{a}.
\end{equation}
If we want to have other types of Bell states, we can implement Hadamard gates of $|\rangle_{b_{1}}$ and 
$|\rangle_{b_{2}}$ respectively on above state, which yields,
$$|g\rangle_{1}|g\rangle_{2}(C|00\rangle-iD|11\rangle)_{b_{1}b_{2}}|0\rangle_{a} \rightarrow $$
\begin{equation}
\frac {1}{2}|g\rangle_{1}|g\rangle_{2}[(C-iD)(|00\rangle+|11\rangle)_{b_{1}b_{2}}-
(C+iD)(|01\rangle+|10\rangle)_{b_{1}b_{2}}]|0\rangle_{a}.
\end{equation}
By choosing values of $C=\pm iD$, we can obtain different Bell states between motional states
of two trapped ions,  which would be useful in future teleportation experiments with massive particles
involved. With the similar idea, we can have entanglement between internal states of the two ions.
Suppose we start from $|e\rangle_{1}(C|0\rangle+D|1\rangle)_{b_{1}}|g\rangle_{2}|0\rangle_{b_{2}}|0\rangle_{a}$
with $|C|^{2}+|D|^{2}=1$, performing $R^{I_{1}}_{6}(\frac {\pi}{2})$, CNOT$_{ab_{2}}$, and 
$R^{I_{2}}_{4}(\frac {\pi}{2})$ sequentially will yield
$$|e\rangle_{1}(C|0\rangle+D|1\rangle)_{b_{1}}|g\rangle_{2}|0\rangle_{b_{2}}|0\rangle_{a}  
\rightarrow (C|e\rangle_{1}|00\rangle_{ab_{1}}-iD|g\rangle_{1}|10\rangle_{ab_{1}})|g\rangle_{2}|0\rangle_{b_{2}}$$
\begin{equation}
\rightarrow C|e\rangle_{1}|0\rangle_{b_{1}}|g\rangle_{2}|00\rangle_{ab_{2}}-iD|g\rangle_{1}|0\rangle_{b_{1}}
|g\rangle_{2}|11\rangle_{ab_{2}}\rightarrow (C|eg\rangle-D|ge\rangle)_{12}|000\rangle_{b_{1}b_{2}a}.
\end{equation}   

\section {Discussion and conclusion}

To some extent, our scheme is similar to the well-known model proposed by Cirac and Zoller [21], in which 
some cold ions, confined in a linear ion trap and radiated by laser beams individually, interact with
each other by coupling to the common vibrational motion. In contrast, in our scheme, if we perform swapping 
operations of $S_{ab}$ alternatively in the two traps, it is easy to obtain, 
$$|g\rangle_{1}|g\rangle_{2}|0\rangle_{b_{1}}|0\rangle_{b_{2}}|0\rangle_{a} 
\rightarrow |g\rangle_{1}|g\rangle_{2}|0\rangle_{b_{1}}|0\rangle_{b_{2}}|0\rangle_{a},$$
$$|g\rangle_{1}|g\rangle_{2}|0\rangle_{b_{1}}|1\rangle_{b_{2}}|0\rangle_{a} 
\rightarrow |g\rangle_{1}|g\rangle_{2}|0\rangle_{b_{1}}|1\rangle_{b_{2}}|0\rangle_{a},$$
$$|g\rangle_{1}|g\rangle_{2}|1\rangle_{b_{1}}|0\rangle_{b_{2}}|0\rangle_{a} 
\rightarrow |g\rangle_{1}|g\rangle_{2}|1\rangle_{b_{1}}|1\rangle_{b_{2}}|0\rangle_{a},$$
$$|g\rangle_{1}|g\rangle_{2}|1\rangle_{b_{1}}|1\rangle_{b_{2}}|0\rangle_{a} 
\rightarrow |g\rangle_{1}|g\rangle_{2}|1\rangle_{b_{1}}|0\rangle_{b_{2}}|0\rangle_{a}.$$
That means, by coupling the cavity state, we can have quantum computing on motional states of trapped ions. 
Compared with Ref.[21], we have an additional degree of freedom, i.e., internal states of the ions for storing
quantum information. This quantum information storage can be easily made by
state transfer from motional states of ions to corresponding internal states. To
the best of our knowledge, our work is the first scheme to make quantum computing in the 
computational space spanned by ions' motional states and cavity state. It would open a new way to perform 
quantum computing on bosonic qubits. Another distinguished character of our scheme is that only two levels of 
each ions are employed in our model, which is important for avoiding 
detrimental effect from the fluctuation of external magnetic field [22]. 

From above results, we know that the phase factor $\phi$ plays very important role in our scheme. 
If ion traps are located at the node of the standing wave of the cavity field, as in Ref.[10], then 
no case 7 would appear, and thereby it is impossible for us to have a general quantum
computing presented above. Moreover, we have to mention that all the entangled states presented in last section 
are prepared without any measurement. This is important because sometimes it is difficult to make measurement
[23], and sometimes we do not want to have state reduction if preparation of entangled states is in 
the course of quantum information processing. Furthermore, we must mention that both  motional
states of ions and the cavity state are more susceptible to decoherence than internal states of ions.
So accomplishment of desired operations should be much faster than that in former schemes with 
internal states of atoms being qubits. Suppose $g=2\pi\times 3\times 10^{7} Hz$ [24], 
$G=2\pi\times 5\times 10^{5} Hz$ [3], $\eta_{c}\approx\eta_{L}=0.2$ and 
$\phi=\frac {\pi}{4}$, the time for accomplishing a CNOT$_{ab}$ gate, a CNOT$_{ba}$ gate, a Hadamard
gate of $|\rangle_{a}$ and a Hadamard gate of $|\rangle_{b}$ are respectively $1.5\times 10^{-7}~sec$, 
$7.8\times 10^{-6}~sec$, $4.2\times 10^{-6}~sec$ and $6.8\times 10^{-6}~sec$. For more clarity, some numerical
results are demonstrated in Fig.1, which shows the variation of implementation time of different quantum 
gates with respect to $g$ and $G$. Experimentally, 
decoherence time for motional state of a cold trapped ion is about $10^{-3}$ $sec$ [25]. However, the decay 
time of the optical cavity is on $microsec$ time scale. It means that, even if we neglect the delay time 
between any two adjacent operations, which is used to change the laser frequency or cavity frequency for obtaining
different interactions, our scheme can not work well with current experimental techniques. Therefore, 
to implement our scheme, more advanced technique is highly demanded to prolongate the decay time of the 
optical cavity, to enlarge the coupling coefficients $g$ and $G$, and to reduce the time delay between any 
two adjacent operations. We noticed some proposals to increase the value of $G$ in so-called strong-excitation regime
of the trapped ion [26], although no experimental reports in this respect have been presented so far. 
We also noticed the latest development of microwave cavity experiment in which the decay time of the cavity 
mode is as long as 0.2 $sec$ [27]. Moreover, to reduce the time delay, we may use different lasers with 
different frequencies, instead of changing the frequency of a single laser. Nevertheless, 
how to reduce the performance time of our scheme to be shorter than the decoherence time of the
cavity would be a big challenge for achieving our scheme experimentally. If we suppose that there is no 
error in our operation, as lifetime of the motional state of the ion is much longer than the performance time
of our quantum gate, decoherence of the cavity state is the only factor to affect the fidelity of our proposed 
gate. As a simple assessment, we present the numerical 
treatment for the fidelity of $CNOT_{ba}$ performance with respect to the decoherence time of cavity state. For
simplicity, we assume that the cavity photon decay happens in the last step concerning $R^{I}_{4}(\pi/2)$. By
means of the method in Ref.[28], we can find that our scheme works very well as long as the
implementation time is half the time of the cavity photon decay. 

In conclusion, we have considered an ideal situation of the ion-trap-cavity system and proposed an 
interesting 
scheme for performing quantum computing and quantum communication in the system with trapped ions 
placed in a high-Q optical cavity. The entangled states including Bell states and GHZ state can be prepared
without state reduction. In principle, this model can be generalized to the case of many trapped ions in the 
same high-Q optical cavity. However, by using current cavity technique, with the increase of 
the cavity size for containing more trapped ions, the
coupling coefficients $g$ would decrease. This is also the challenge for achieving some quantum computing
schemes with semiconductor quantum dots [29]. Nevertheless, once our scheme is achieved, as the traps  
interact with each other through coupling to the same cavity state, we can have a network of quantum 
information processing. If the decaying effect of both motional states of the ion and the cavity state 
are considered and suitably used, the treatment would be more complicated, but more interesting. In 
particular, if cavities can be connected with each other by quantum wires, the system under consideration 
here would be a workable node of a larger 
quantum network. We have noticed the first experimental report in the ion-trap-cavity system
with single atomic quantum bit confined [30]. With rapid development of experimental technique, we believe
that our proposal would be of interest in exploration of quantum information processing.

The authors would like to thank Irene D'Amico and Paolo Zanardi for valuable discussion.
The work is partly supported by the European Commission through the Research Project SQID within FET Program,
and partly by the National Natural Science Foundation of China. 

$Note$ $added$: after finishing this work, we become aware of a recent work \cite{solano}, in which
maximally entangled states is generated between vibrational states of two trapped ions by illuminating the 
two ions simultaneously with two dispersive Raman pulses. That work is different
from ours because the entanglement between vibrational states of trapped ions is generated in our scheme 
by the connection of cavity state.

\begin{center} {\bf The Captions of the figures}\end{center}

Fig.1 ~~ Time of quantum gate performance with respect to the coupling strength, where $\phi=\pi/4$, 
$\eta_c=\eta_L=0.2$.
(A) Setting $G=2\pi\times 5\times10^{5} Hz$, implementation time varies with values of $g$ (in the unit of
$2\pi\times 30 MHz$). The dotted, dashed and solid curves denote $CNOT_{ba}$, $H_{a}$ and $CNOT_{ab}$
respectively. (B) Setting $g=2\pi\times 3\times 10^{7} Hz$, implementation time corresponds to different G (in
the unit of $\pi$ MHz). The dotted, dashed and solid curves denote $CNOT_{ba}$, $H_{b}$ and $H_{a}$ 
respectively. 

Fig.2 ~~ Fidelity of the $CNOT_{ba}$ gate performance with respect to $T_{im}/T_{d}$, where $T_{im}$ and 
$T_{d}$ are respectively implementation time of $CNOT_{ba}$ gate and decoherence time of the cavity state. 
For cases of Hadamard gates and $CNOT_{ab}$, we can have similar results.

\begin{thebibliography}{27}
\bibitem{23} For example, in Eq.(15), if states of phonon and photon are entangled with internal states of the
two ions, it is impossible to obtain the entangled internal states of the two ions by making measurement on
phonon and photon states.
\end{thebibliography}
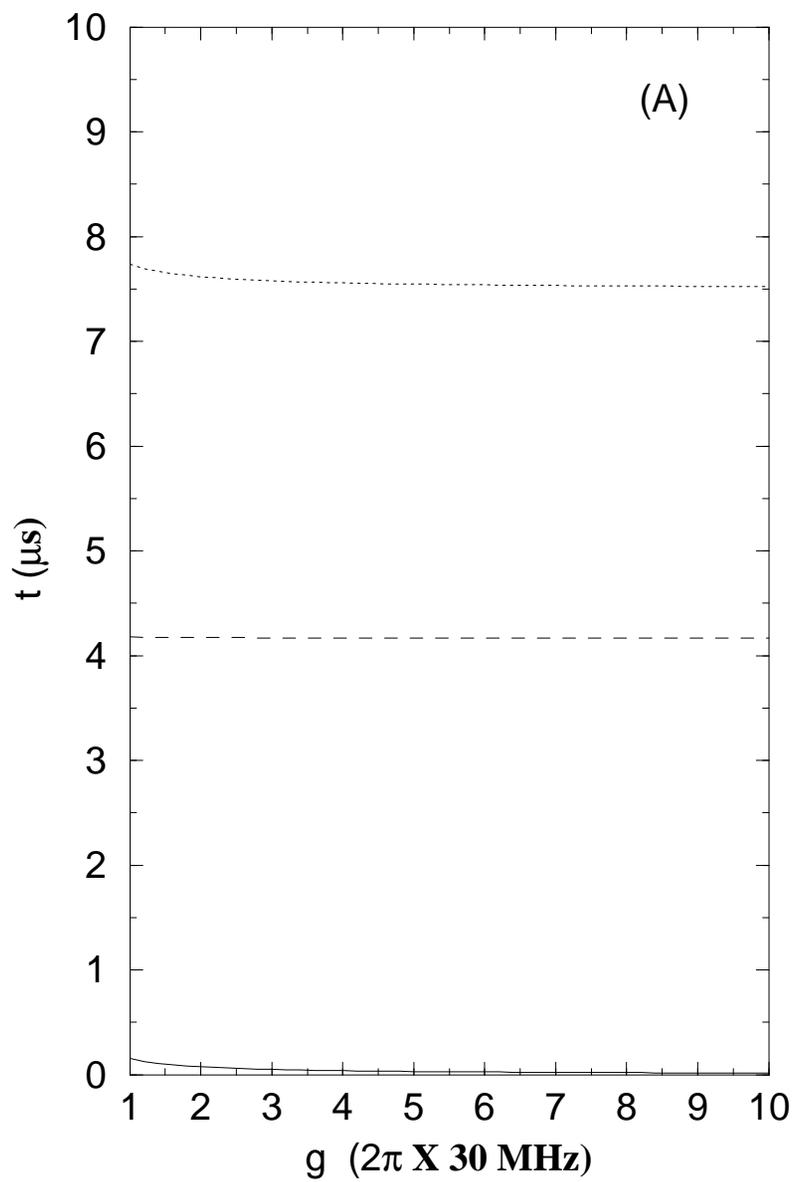
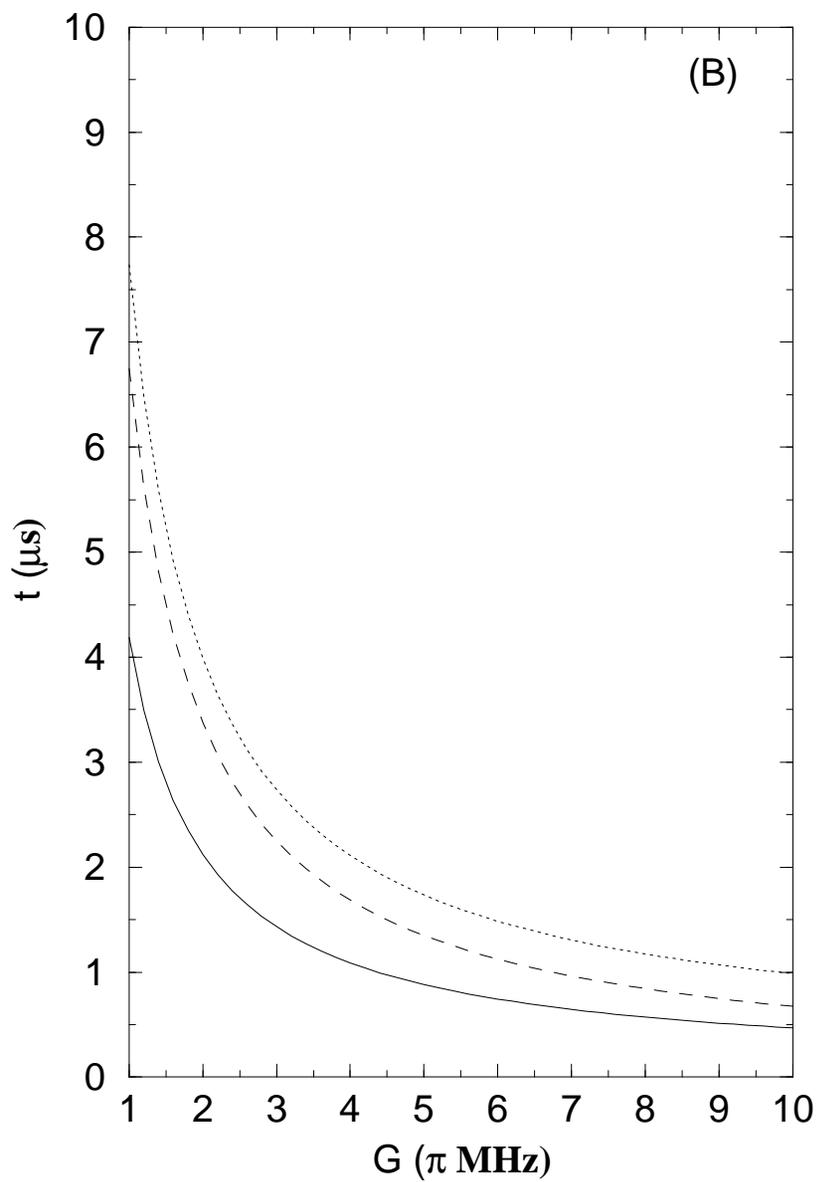
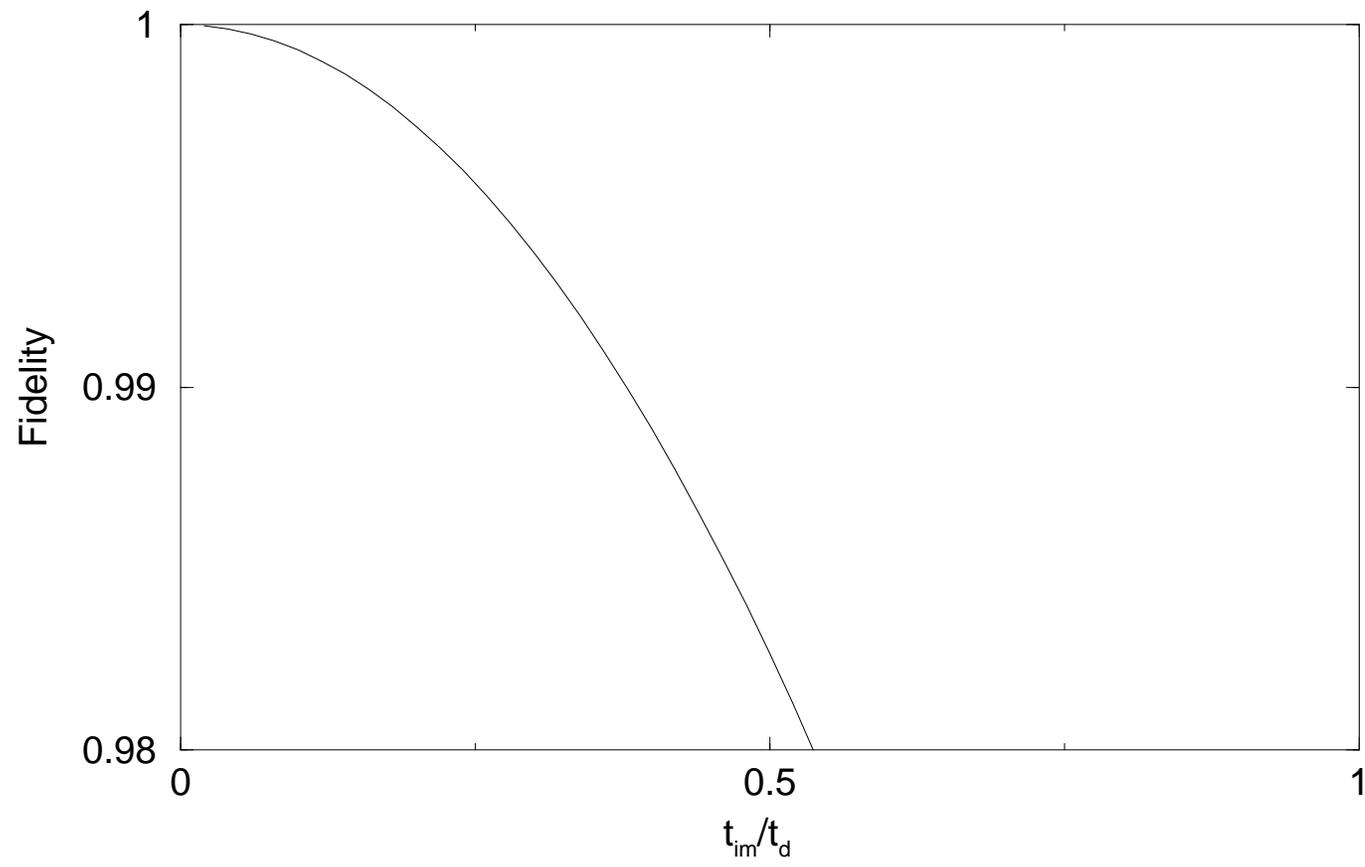
\end{document}